# New constraints on Saturn's interior from Cassini astrometric data

Valéry Lainey[1], Robert A. Jacobson[2], Radwan Tajeddine[3,1], Nicholas J. Cooper[4,1], Carl Murray[4], Vincent Robert[5,1], Gabriel Tobie[6], Tristan Guillot[7], Stéphane Mathis[8], Françoise Remus[9,1,8], Josselin Desmars[10,1], Jean-Eudes Arlot[1], Jean-Pierre De Cuyper[11], Véronique Dehant[11], Dan Pascu[12], William Thuillot[1], Christophe Le Poncin-Lafitte[13], Jean-Paul Zahn[9,†]

*[1]IMCCE, Observatoire de Paris, PSL Research University, CNRS-UMR8028 du CNRS, UPMC, Lille-1, 77 Av. Denfert-Rochereau, 75014, Paris, France*

*[2]Jet Propulsion Laboratory, California Institute of Technology, 4800 Oak Grove Drive Pasadena, California 91109-8099*

*[3] Center for Radiophysics and Space Research, Cornell University, 326 Space Sciences Building, Ithaca, NY 14853*

*[4]Queen Mary University of London, Mile End Rd, London E1 4NS, United Kingdom*

*[5]IPSA, 7-9 rue Maurice Grandcoing, 94200 Ivry-sur-Seine, France*

*[6]Laboratoire de Planétologie et Géodynamique de Nantes, Université de Nantes, CNRS, UMR 6112, 2 rue de la Houssinière, 44322 Nantes Cedex 3, France*

*[7]Laboratoire Lagrange, CNRS UMR 7293, Université de Nice-Sophia Antipolis, Observatoire de la Côte d'Azur, B.P. 4229 06304 Nice Cedex 4, France*

*[8]Laboratoire AIM Paris-Saclay, CEA/DSM - Université Paris Diderot - CNRS, IRFU/SAp Centre de Saclay, 91191 Gif-sur-Yvette, France*

*[9]LUTH-Observatoire de Paris, UMR 8102 du CNRS, 5 place Jules Janssen, 92195 Meudon Cedex, France*

*[10]Observatório Nacional, Rua José Cristino 77, São Cristovão, Rio de Janeiro CEP 20.921-400, Brazil*

*[11]Royal Observatory of Belgium, Avenue Circulaire 3, 1180 Uccle, Bruxelles, Belgium*

*[12]USNO (retired), 3450 Massachusetts Avenue Northwest, Washington, DC 20392, United States*

*[13]SYRTE, Observatoire de Paris, PSL Research University, CNRS, Sorbonne Universités, UPMC Univ. Paris 06, LNE, 61 avenue de l'Observatoire, 75014 Paris, France*

**Corresponding author:** V.Lainey (lainey@imcce.fr)

**Abstract**

Using astrometric observations spanning more than a century and including a large set of Cassini data, we determine Saturn's tidal parameters through their current effects on the orbits of the eight main and four coorbital moons. We have used the latter to make the first determination of Saturn's Love number, $k_2$=0.390 ± 0.024, a value larger than the commonly used theoretical value of 0.341 (Gavrilov & Zharkov, 1977), but compatible with more recent models (Helled & Guillot, 2013) for which $k_2$ ranges from 0.355 to 0.382. Depending on the assumed spin for Saturn's interior, the new constraint can lead to a reduction of up to 80% in the number of potential models, offering great opportunities to probe the planet's interior. In addition, significant tidal dissipation within Saturn is confirmed (Lainey et al., 2012) corresponding to a high present-day tidal ratio $k_2/Q=(1.59 \pm 0.74) \times 10^{-4}$ and implying fast orbital expansions of the moons. This high dissipation, with no obvious variations for tidal frequencies corresponding to those of Enceladus and Dione, may be explained by viscous friction in a solid core, implying a core viscosity typically ranging between $10^{14}$ and $10^{16}$ Pa.s (Remus et al., 2012). However, a dissipation increase by one order of magnitude at Rhea's frequency could suggest the existence of an additional, frequency-dependent, dissipation process, possibly from turbulent friction acting on tidal waves in the fluid envelope of Saturn (Ogilvie & Lin, 2004). Alternatively, a few of Saturn's moons might themselves experience large tidal dissipation.

## 1 Introduction

Tidal effects among planetary systems are the main driver in the orbital migration of natural satellites. They result from physical processes arising in the interior of celestial bodies, not observable necessarily from surface imaging. Hence, monitoring the moons' motions offers a unique opportunity to probe the interior properties of a planet and its satellites. In common with the Martian and Jovian systems (Lainey et al., 2007, 2009), the orbital evolution of the Saturnian system due to tidal dissipation can be derived from astrometric observations of the satellites over an extended time period. In that respect, the presence of the Cassini spacecraft in orbit around Saturn since 2004 has provided unprecedented astrometric and radio-science data for this system with exquisite precision. These data open the door for estimating a potentially large number of physical parameters simultaneously, such as the gravity field of the whole system and even separating the usually strongly correlated tidal parameters $k_2$ and Q.

The present work is based on two fully independent analyses (modelling, data, fitting procedure) performed at IMCCE and JPL, respectively. Methods are briefly described in Section 2. Section 3 provides a comparison between both analyses as well as a global solution for the tidal parameters $k_2$ and Q of Saturn. Section 4 describes possible interior models of Saturn compatible with our observations. Section 5 discusses possible implications associated with the strong tidal dissipation we determined.

## 2. Material and methods

Both analyses stand on numerical computation of the moons' orbital states at any time, as well as computation of the derivatives of these state vectors with respect to: i) their initial state for some reference epoch; ii) many physical parameters. Tidal effects between both the moons and the primary are introduced by means of the two classical quantities $k_2$ and Q. We recall that the so-called Love number $k_2$ describes the response of the potential of the distorted body experiencing tides. Q, often called the quality factor (Kaula 1964), is inversely proportional to the amount of energy dissipated essentially as heat by tidal friction. Coupled tidal effects such as tidal bulges raised on Saturn by one moon and acting on another are considered. Besides the eight main moons of Saturn, the coorbital moons Calypso, Telesto, Polydeuces, and Helene are integrated in both studies.

Although the two tidal parameters $k_2$ and Q often appear independently in the equations of motion, the major dynamical effect by far is obtained when the tide raised by a moon on its primary acts back on this same moon. In this case, only the ratio $k_2/Q$ is present as a factor for the major term, therefore preventing an independent fit of $k_2$ and Q. However, the small co-orbital satellites raise negligible tides on Saturn and yet react to the tides raised on the planet by their parent satellites. This unique property allows us to make a fit for $k_2$ that is almost independent of Q (see Appendix A1). In particular, we find that the modelling of such cross effects between the coorbital moons allows us to obtain a linear correlation between $k_2$ and Q of only 0.03 (Section 3 and Appendix A4). Thanks to the inclusion of Telesto, Calypso, Helene and Polydeuces, we can estimate $k_2$ essentially around the tidal frequencies of Tethys and Dione.

## 2.1 IMCCE's approach

The IMCCE approach benefits from the NOE numerical code that was successfully applied to the Mars, Jupiter, and Uranus systems (Lainey et al., 2007, 2008, 2009). It integrates the full equations of motion for the centre of mass of the satellites and solves for the partial derivatives of the system. This latter set of equations allows for a fitting procedure to the observations. For a complete description of the equations solved, we refer to Lainey et al. (2012) and references therein.

Here, fourteen moons of Saturn are considered all together, i.e. the eight main moons and six coorbital moons (Epimetheus, Janus, Calypso, Telesto, Helene, and Polydeuces). All the astrometric observations already considered in Lainey et al. (2012) and Desmars et al. (2009) are used, with the addition of a large set of ISS-Cassini data (Tajeddine et al., 2013, 2015; Cooper et al. 2014). We also include a new reduction of old photographic plates, obtained at USNO between the years 1974 and 1998. As part of the ESPaCE European project, the scanning and new astrometric reduction of these plates were performed recently at Royal Observatory of Belgium and IMCCE, respectively (Robert et al. 2011; to be submitted). We use a weighted least squares inversion procedure and minimize the squared differences between the observed and computed positions of the satellites in order to determine the parameters of the model. For each fit, the following parameters are released simultaneously and without constraints: the initial state vector and mass of each moon, the mass, the gravitational harmonic $J_2$, the orientation and the precession of the pole of Saturn as well as its tidal parameters $k_2$ and Q. No da/dt term is released for Mimas. In particular, it appears that the large signal obtained in Lainey et al. (2012) can be removed after fitting the gravity field of the Saturn system.

## 2.2 JPL's approach

The second approach incorporates the tidal parameters into the ongoing determination of the satellite ephemerides and Saturnian system gravity parameters that support navigation for the Cassini Mission. Initial results from that work appear in Jacobson et al. (2006). For Cassini the satellite system is restricted to the eight major satellites, Phoebe, and the Lagrangians Helene, Telesto, and Calypso. The analysis procedure is to repeat all of the Cassini navigation reconstructions but with a common set of ephemerides and gravity parameters. We combine these new reconstructions with other non-Cassini data sets to obtain the updated ephemerides and revised gravity parameters. The non-Cassini data include radiometric tracking of the Pioneer and Voyager spacecraft, imaging from Voyager, Earth-based and HST astrometry, satellite mutual events (eclipses and occultations), and Saturn ring occultations. We process the data via a weighted least-squares fit that adjusts our models of the orbits of the satellites and the four spacecraft (Pioneer, Voyager 1, Voyager 2, Cassini). Peters (1981) and Moyer (2000) describe the orbital models for the satellites and spacecraft, respectively. The set of gravity related parameters adjusted in the fit contains the GMs of the Saturnian system and the satellites (Helene, Telesto, and Calypso are assumed massless), the gravitational harmonics of Saturn, Enceladus, Dione, Rhea, and Titan, Saturn's polar moment of inertia, the orientation of Saturn's pole, and the tidal parameters $k_2$ and Q.

## 3. Results

Since tidal effects within Saturn and Enceladus have almost opposite orbital consequences, Lainey et al. (2012) could not solve for the Enceladus tidal ratio $k_2^E/Q^E$. Here, we face a similar strong correlation and follow their approach by considering two extreme scenarios for Enceladus' tidal state. In a first inversion, we neglect dissipation in Enceladus and obtain for Saturn $k_2$, $k_2^{(I)}=0.371 \pm 0.003$, $k_2^{(J)}=0.381 \pm 0.011$ (formal error bar, $1\sigma$) where the indices *I* and *J* refer to the IMCCE and JPL solutions, respectively. The Saturn tidal ratio that we obtain is $k_2/Q^{(I)}=(1.32 \pm 0.25) \times 10^{-4}$, $k_2/Q^{(J)}=(1.04 \pm 0.19) \times 10^{-4}$). In a second inversion, we assume Enceladus to be in a state of tidal equilibrium (Meyer & Wisdom, 2007), obtaining $k_2^{(I)}=0.372 \pm 0.003$, $k_2^{(J)}=0.402 \pm 0.011$ and $k_2/Q^{(I)}=(2.07 \pm 0.26) \times 10^{-4}$, $k_2/Q^{(J)}=(1.22 \pm 0.23) \times 10^{-4}$. If both studies are generally in good agreement within the uncertainty of the measurements (Extended Data), the last $k_2/Q^{(I)}$ value stands at 3σ of the JPL estimation. This possibly reflects the difference in the data sets, since JPL introduced radio-science data, while IMCCE introduced scanning data. Nevertheless, both estimates suggest strong tidal dissipation, at least about five times larger than previous theoretical estimate (Sinclair, 1983). Merging IMCCE's and JPL's results into one value by overlapping the extreme 1σ values, we get $k_2$=0.391 ± 0.023 and $k_2/Q$=(1.59 ± 0.74) × $10^{-4}$. These last error bars are not formal 1σ values anymore, but the likely interval of expected physical values.

Last, to assess a possibly large variation of Saturn Q as function of tidal frequency, we followed Lainey et al. (2012) and released as free parameters four different Saturnian tidal ratios $k_2/Q$ associated with the Enceladus', Tethys', Dione's, and Rhea's tides (see Tables 1-2). It turns out that no significant change for the $k_2$ estimation arises with an overall result of $k_2$=0.390 ± 0.024. Moreover, global solutions for $k_2/Q$ ratios are equal to (20.70 +/- 19.91) x $10^{-5}$, (15.84 +/- 12.26) x $10^{-5}$, (16.02 +/- 12.72) x $10^{-5}$, (123.94 +/- 17.27) x $10^{-5}$ at Enceladus', Tethys', Dione's and Rhea's tidal frequency, respectively. We provide in Figure 1 a plot showing all global $k_2/Q$ ratios associated with constant and non-constant assumptions.

| | $k_2$ | $k_2/Q$ (S2) | $k_2/Q$ (S3) | $k_2/Q$ (S4) | $k_2/Q$ (S5) |
|---|---|---|---|---|---|
| IMCCE | 0.372 +/- 0.003 | (7.4 +/- 3.1) x $10^{-5}$ | (10.9 +/- 6.1) x $10^{-5}$ | (16.1 +/- 3.8) x $10^{-5}$ | (122.3 +/- 15.0) x $10^{-5}$ |
| JPL | 0.377 +/- 0.011 | (5.5 +/- 4.7) x $10^{-5}$ | (6.0 +/- 2.4) x $10^{-5}$ | (21.5 +/- 7.3) x $10^{-5}$ | (125.8 +/- 14.9) x $10^{-5}$ |

Table 1: *Fitting $k_2$ and variable Saturnian Q at S2..S5 frequencies.*

| | $k_2$ | $k_2/Q$ (S2) | $k_2/Q$ (S3) | $k_2/Q$ (S4) | $k_2/Q$ (S5) |
|---|---|---|---|---|---|
| IMCCE | 0.372 +/- 0.003 | (18.1 +/- 3.1) x $10^{-5}$ | (11.9 +/- 6.1) x $10^{-5}$ | (15.0 +/- 3.8) x $10^{-5}$ | (121.6 +/- 15.0) x $10^{-5}$ |
| JPL | 0.394 +/- 0.011 | (27.1 +/- 13.5) x $10^{-5}$ | (21.5 +/- 6.6) x $10^{-5}$ | (5.4 +/- 2.1) x $10^{-5}$ | (127.9 +/- 13.3) x $10^{-5}$ |

Table 2: *Fitting $k_2$ and variable Saturnian Q at S2..S5 frequencies assuming Enceladus' tidal equilibrium.*

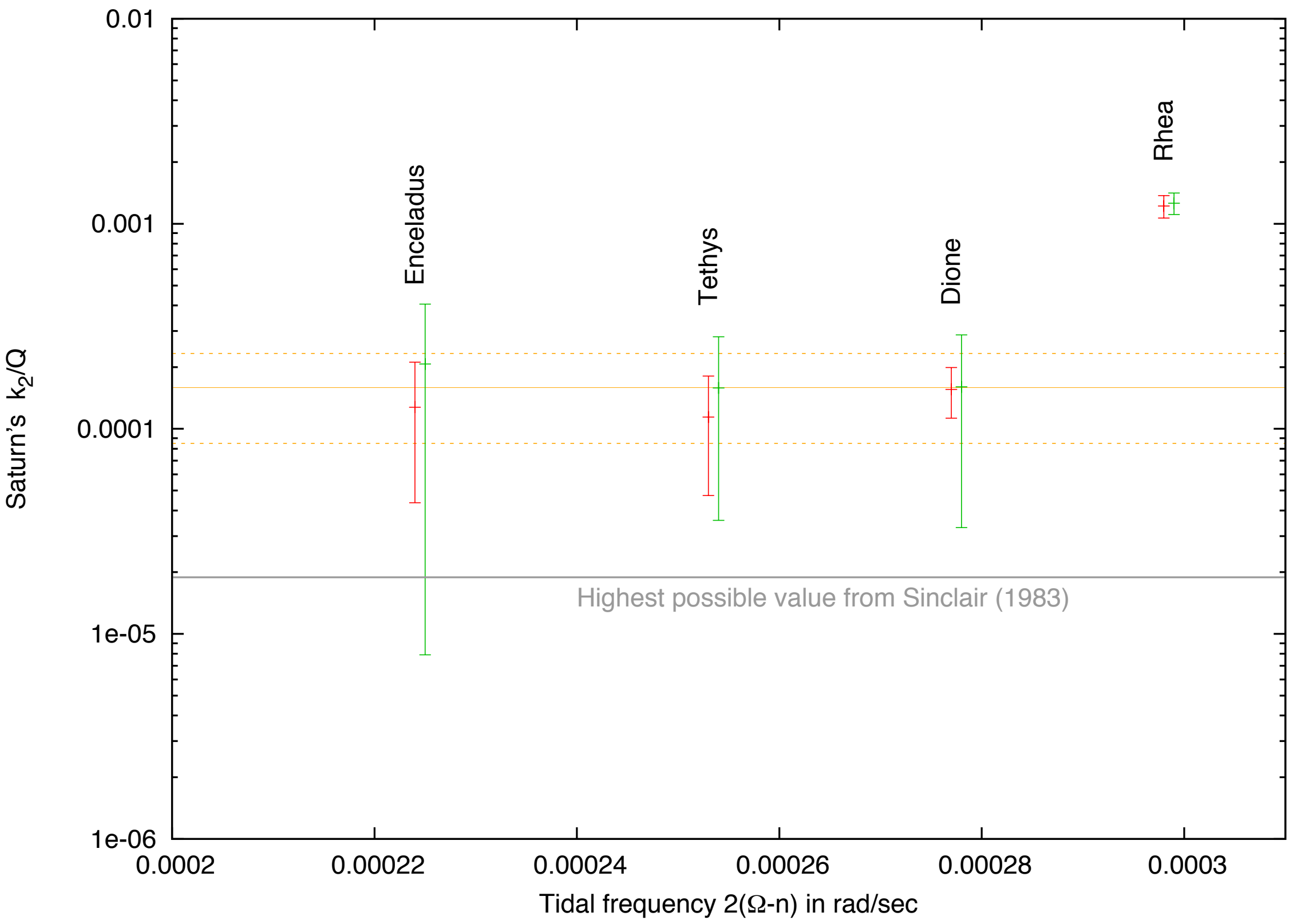

Figure 1: *Variation of the Saturnian tidal ratio $k_2/Q$ as a function of tidal frequency $2(\Omega\text{-}n)$, where $\Omega$ and $n$ denote its rotation rate and the moon's mean motion, respectively. Four frequencies are presented associated with Enceladus', Tethys', Dione's and Rhea's tides. IMCCE and JPL solutions are in red and green, respectively. They are shown slightly shifted from each other along the X-axis for better visibility. Orange lines refer to the global estimation $k_2/Q = (15.9 \pm 7.4) \times 10^{-5}$.*

## 4. Modeling Saturn's interior

To model the tidal response of Saturn's interior and to compare it to the $k_2$ and $Q$ values inferred in the present study, we consider a wide range of interior models consistent with the gravitational coefficients measured using the Cassini spacecraft (Helled & Guillot 2013). In total, 302 interior models, corresponding to various core size and composition, helium phase separation and enrichment in heavy elements in the external envelope, have been tested. Each interior model is characterized by radial profiles of density, $r$, and bulk modulus, $K$.

The tidal response of Saturn's interior is computed from all the considered density profiles assuming that the core is solid and viscoelastic, with radius $R_{core}$ (varying typically between 7000 and 16000 km) overlaid by a thick non-dissipative fluid envelope, similar to the approach of Remus et al. (2012, 2015). The Love number $k_2$ and the global dissipation function $Q^{-1}$ are determined by integrating the 5 radial functions, $y_i$, describing the displacements, stresses, and gravitational potential from the planet center to the surface. The viscoelastic deformation in the solid viscoelastic core is computed using the compressible elastic formulation of Takeuchi & Saito (1972), adapted to viscoelastic media (see Tobie et al., 2005 for more details). For the fluid envelope, the static formulation of Saito (1974) is used. The system of differential equations (6 in the core and 2 in the envelope) is solved by integrating from the center to the surface three independent solutions using a fifth order Runge-Kutta method with adaptive stepsize control, and by applying the appropriate condition at the solid core/fluid envelop interface and at the surface (see Takeushi & Saito 1972 and Tobie et al. 2005 for more details). The complex Love number $k_2^c$ is determined from the complex 5$^{th}$ radial function at the planet surface, $y_5^c(R_s)$, and the global dissipation function by the ratio between the imaginary part and the module of $k_2^c$:

$k_2=|k_2^c|=|y_5^c(R_s)-1|;\ Q^{-1}=Im(k_2^c)/|k_2^c|$.

For the solid core, a compressible Maxwell rheology, characterized by the bulk modulus $K$, the shear modulus $\mu$, and the viscosity $\eta$, is assumed. The shear modulus is determined from the bulk modulus assuming a constant $\mu/K$ ratio varying between 0.001 and 1, and the viscosity is assumed constant over a range varying between $10^{12}$ and $10^{18}$ Pa.s.

In order to test the validity of our numerical code, we compared our numerical solutions with the analytical solutions derived by Remus et al. (2012) for a viscoelastic core and a fluid envelope with constant density. As illustrated on Figure 2, we reproduce almost perfectly the analytical value of the tidal Love number. For the dissipation function, the agreement is also very good, the difference between the analytical and numerical solutions never exceed a few per cent. To further test our code, we also compared with the solution provided by Kramm et al. (2011) for a density distribution of a $n=1$ polytrope: we obtained $k_2$=0.5239, while the value reported by Kramm et al. (2011) is 0.5198, which corresponds to a difference of less than 0.8%.

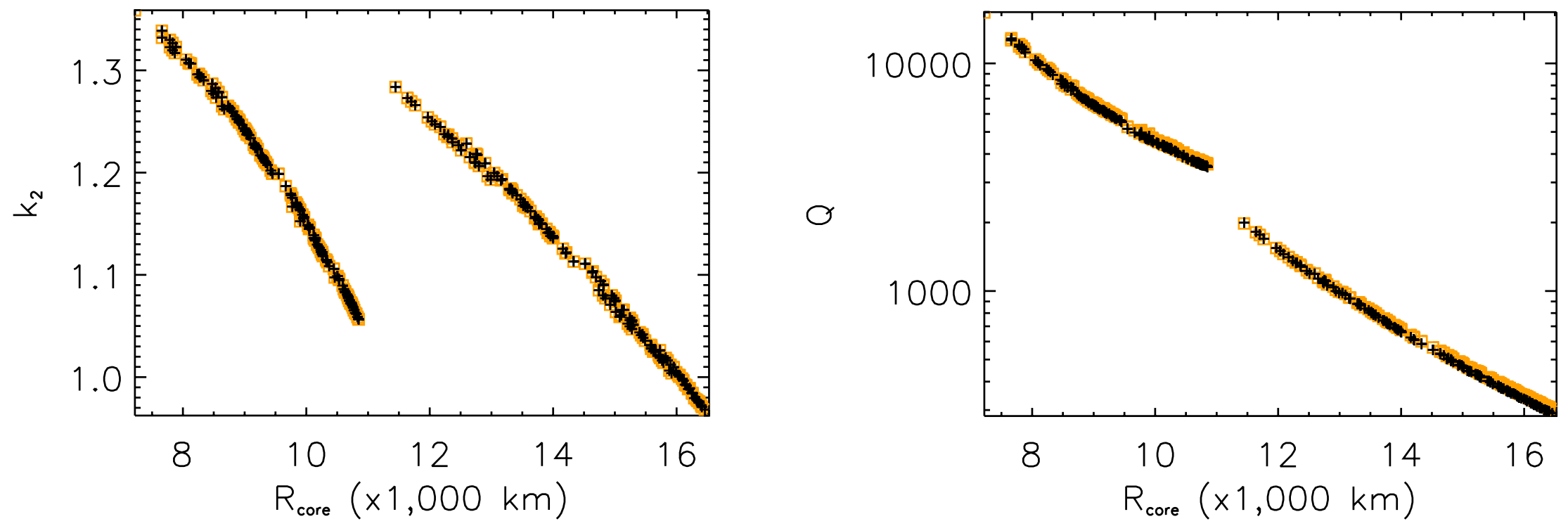

Figure 2: *Comparison between numerical (black crosses) and analytical (orange squares) solutions of tidal Love number, $k_2$ (left) and dissipation factor, Q (right) as a function of core radius, $R_{core}$, computed for a solid viscoelastic core and a fluid envelop with constant density, assuming a core viscosity of $10^{15}$ Pa.s and a shear modulus of 1000 GPa.*

Our calculations confirm that the tidal Love number of the planet is almost entirely determined by the density profile; therefore it is very close to the fluid Love number. The mechanical properties of the core have only very minor influence on the amplitude of $k_2$; they mostly affect the imaginary part of $k_{2c}$, and hence the dissipation factor, *Q*. As shown on Figure 3, the global *Q* factor depends on the assumed shear modulus (hence the $\mu/K$ ratio) and the viscosity in the core as well as on its size. The *Q* factor decreases with increasing core radius and shear modulus. For the largest core radii and $\mu/K$~0.1-0.5, consisting of an ice core, *Q* values lower than 200-300 can be obtained, and *Q* remains below 3000 for viscosity values ranging between about $2.10^{13}$ and $2.10^{16}$ Pa.s. For small core radii (< 11,000 km) corresponding to a rocky core, Q values lower than 3000 can also be found, but for a more restricted range of viscosity values, between typically $10^{15}$ and $10^{16}$ Pa.s. For a very low $\mu/K$ ratio (0.01), Q< 3000 can be obtained for large ice-rich cores and viscosity values of the order of $5.10^{13}$-$5.10^{14}$ Pa.s. These possible ranges of viscosity are compatible with those derived previously in Remus et al. (2012, 2015) where simplified two-layer planetary models were used. As illustrated in Figure 4, the computed $k_2/Q$ values are only weakly sensitive to the tidal frequency. Therefore, even though Q values as low as 200 can be obtained for large cores and appropriate viscoelastic parameters, it is not possible to explain with viscoelastic dissipation Q values of the order of a few thousands at Enceladus' tidal frequency and of a few hundred at Rhea's tidal frequency. Additional dissipation processes in the gaseous deep envelope are thus required to explain the high dissipation inferred from observation at Rhea's tidal frequency (Ogilvie & Lin 2004).

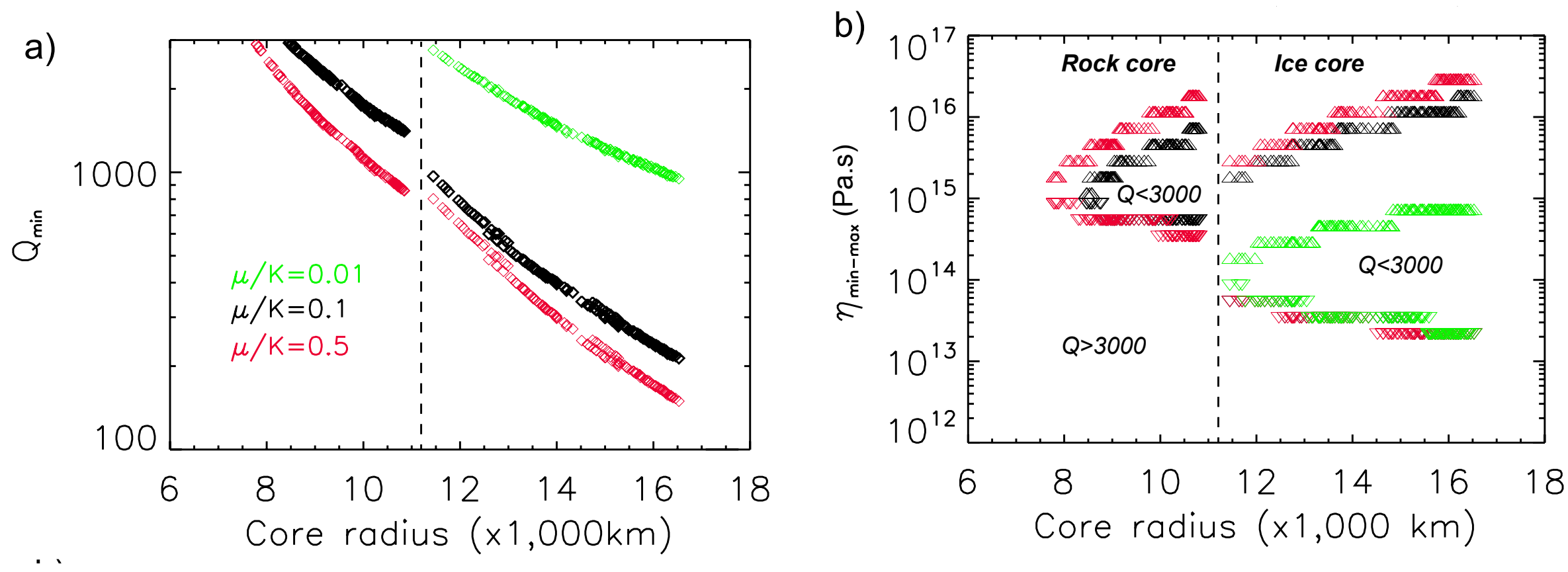

Figure 3: *a) minimum value of the dissipation factor, $Q_{min}$, as a function of core radius for three different values of μ/K (0.01, 0.1, 0.5); (b) Range of viscosity values, $\eta_{max}(\Delta)$ -$\eta_{min}$ ($\nabla$), for which Q<3000 for the three μ/K ratios displayed in (a). The dashed line indicates the transition between high density (rock-dominated) core and low density (ice-dominated) core. For this computation, the tidal frequency was fixed at 2.6 x$10^{-4}$ rad.$s^{-1}$*

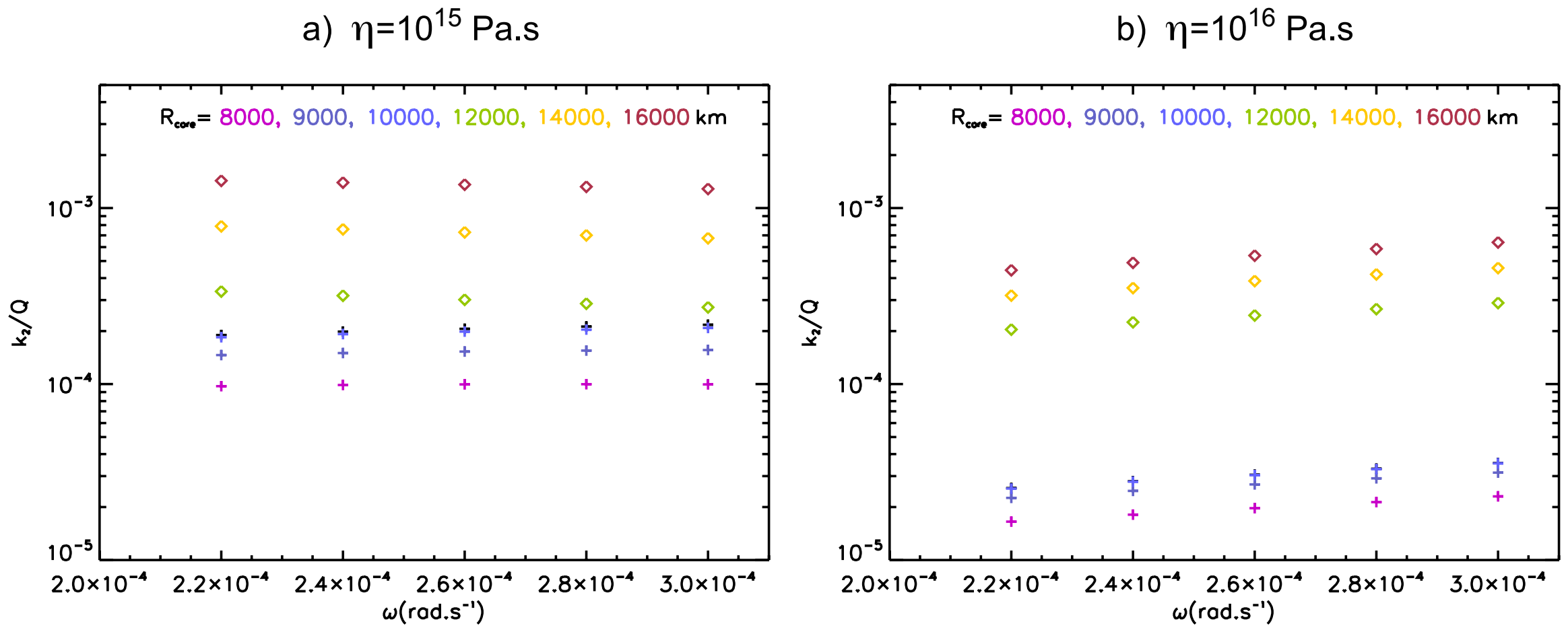

Figure 4: *$k_2/Q$ values as a function of tidal frequency, w, for two core viscosity values ($10^{15}$ (a) and $10^{16}$ (b) Pa.s) for six different values of core radius. The μ/K ratio was fixed to 0.1 for these calculations.*

## 5. Discussion

In 1977, Gavrilov and Zharkov (1977) computed the value of Saturn's Love numbers and obtained for the lowest degree quadripolar coefficient $k_2$=0.341. Even though this value is often used as the reference, it stands on physical assumptions and internal structure models that have since been improved (Guillot 1999, 2005; Hubbard et al., 2009; Kramm et al., 2011; Nettelmann et al., 2013; Helled & Guillot, 2013). Using the models of Helled and Guillot (2013), we show in Figure 2 that for these three-layer models including the uncertainty of differential rotation in the interior gives values of $k_2$ that range between 0.355 and 0.381. About 23% of these models are incompatible with our determination of $k_2$. When focusing on models with a dense core (i.e. in effect using an EOS for pure rocks for the central core), this fraction increases to 47%. It becomes 84% for interior models compatible with the latest estimate of Saturn's spin (Helled et al., 2015), i.e., only 4 models then satisfy the available constraints. All of them have a low-density core (modelled with the EOS of pure ice) and a helium separation occurring at 1 Mbar, in line with recent determinations of hydrogen-helium phase separation (Morales et al., 2009). Understanding more precisely the consequences for our knowledge of Saturn's interior will require dedicated models, but this clearly shows the great potential of the method and its complementarity to studies based on the determination of the planet's gravity field. Any further improvement in the estimation of $k_2$ and the spin rate will allow even better constraints on Saturn's interior.

Our estimation of Saturn's Q confirms the values previously derived by Lainey et al. (2012), which is one order of magnitude smaller than the value derived from the usually expected long term evolution of the moons over the age of the Solar system (Sinclair, 1983). Such low Q or high dissipation rate, implying rapid orbital expansion, suggests that either the dissipation has significantly changed over time, or that the moons formed later after the formation of the Solar system (Charnoz et al. 2011; Ćuk 2014). Since tidal dissipation may arise both in the planet's fluid envelope and its presumably solid core (Guenel et al. , 2014), we can look in more detail at the frequency dependency of the tidal ratio $k_2/Q$ showed in Figure 1. Despite large error bars, the tidal ratios associated with Enceladus, Tethys and Dione do not depart from their former constant estimate. On the other hand, we obtain a strong increase of dissipation at Rhea's frequency. Such a dissipation corresponds to an orbital shift in the longitude of about 75 km (see Appendix A2). The fact that the strong orbital shift at Rhea is observed using both the IMCCE and JPL models, makes systematic errors unlikely. As Rhea has no orbital resonance with any other moon, and no significant dynamical interaction with the rings, its strong orbital shift is more likely the consequence of strong tides.

The rather constant dissipation inferred at tidal frequencies associated with Enceladus, Tethys and Dione suggests dissipation processes dominated by anelastic tidal friction in a solid core (Remus et al., 2012, 2015). In order to test this hypothesis further, we computed the tidal dissipation factor, Q, for the set of internal models presented in Figure 5 and by considering the wide range of viscosity and shear modulus values for the solid core presented in Section 4.

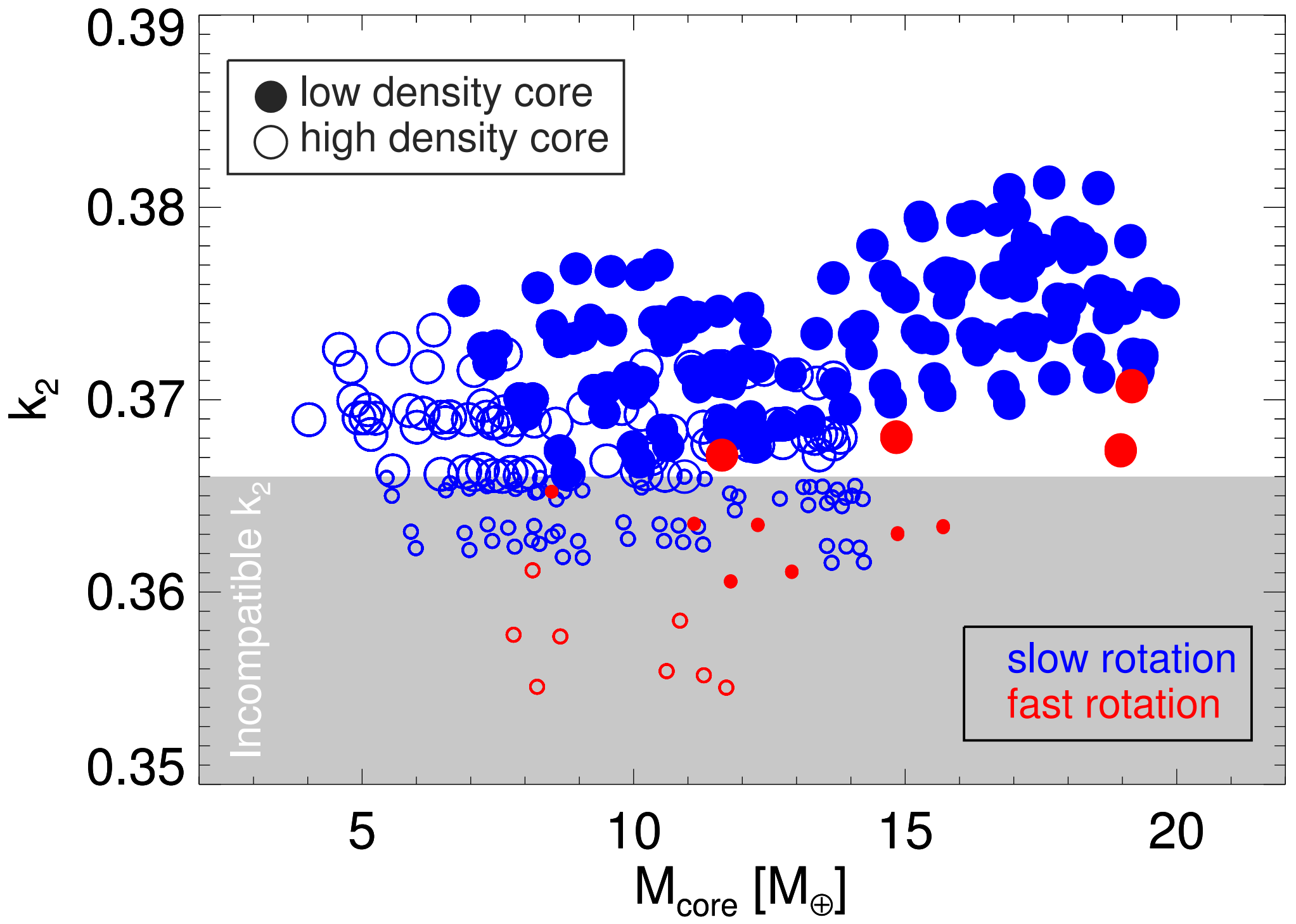

Figure 5: *Mass of the core and $k_2$ Love number for interior models of Saturn from Helled & Guillot (2013). Filled circles indicate models assuming a low density core (modelled using the equation of state of pure ice) while empty circles indicate models assuming a high density core (modelled using the EOS of rocks). Models in blue assume a "slow" deep rotation of 10h39m while models in red assume a "fast" deep rotation of 10h32m, more in line with the recent determination of Helled et al. (2015). The grey area indicates where values of $k_2$ are incompatible with our astrometric determination.*

We showed that for all interior models consistent with $k_2$>0.37, a Q factor lower than 3000 can be obtained for viscosity values ranging typically between $10^{13}$-$10^{16}$ Pa.s for a low density core (Rcore > 11,000 km) and for a more restricted viscosity range ($10^{15}$-$10^{16}$ Pa.s) for a high density core (Rcore < 11,000 km). For the largest core radii and $\mu/K$~0.1-0.5, Q values lower than 200-300, compatible with Rhea's estimate, can be obtained. However, a Q factor compatible with the innermost moons and Rhea simultaneously cannot be found, as the viscoelastic solution is only weakly frequency-dependent (see Figure 4 and Remus et al. (2012)). Excluding significant tidal dissipation in moons other than Enceladus, additional tidal friction processes are needed to explain the smaller Q factor at Rhea's frequency. The best candidate is turbulent friction applied to tidal inertial waves (their restoring force is the Coriolis acceleration) in the deep, rapidly rotating, oblate convective envelope of Saturn that dissipates their kinetic energy (Ogilvie & Lin, 2004; Braviner & Ogilvie, 2015). This fluid dissipation is resonant and its amplitude can therefore vary by several orders of magnitude as a function of the tidal frequency and of the effective turbulent viscosity (Ogilvie & Lin, 2004). Hence, it can explain the increase by one order of magnitude of the dissipation over the small frequency range arising between Dione and Rhea.

A more speculative explanation might be that Saturn's tidal dissipation essentially occurs in the core, but that several other moons, in addition to Enceladus, themselves experience large tidal dissipation. Since this latter effect has opposite orbital consequences to tides in the primary, orbital expansion could show up at moderate levels for most studied frequencies, despite a potential low Q solution for Saturn. Such a hypothesis could be consistent with a possible global ocean under Mimas (Tajeddine et al., 2014). Interestingly, this would provide an increase of Titan's eccentricity over time, partly explaining its current high value (see Appendix A3). Extending the astrometric study to more Saturnian moons or measuring the moons' obliquity will help test such a hypothesis.

## 5. Conclusion

Using a large set of astrometric observations including ground observations and thousands of ISS-Cassini data, we provide the first estimation of the Love number of Saturn $k_2$. Moreover, we confirm the strong tidal dissipation found by Lainey et al. (2012), but associated with an intense frequency-dependent peak of tidal dissipation for Rhea's tidal frequency. Modelling the likely interior of Saturn, it appears two different tidal mechanisms may arise within the planet. The first one is the tidal friction within the dense core of the primary, while significant tidal dissipation may occur inside the outer envelope at Rhea's tidal frequency. Nevertheless, we cannot rule out a second scenario, which considers tidal dissipation within Saturn's core, only. In that case, significant tidal dissipation inside moons other than Enceladus shall occur.

## Appendix

### A1 - The tidal effects on coorbital satellites

The effects of tidal bulges on one moon's motion are generally far below detection, unless those tides are raised by the same moon. Indeed, such a configuration produces a secular effect on the orbit that may be detectable after a sufficient amount of time. On the other hand, tidal bulges associated with another moon will introduce essentially quasi-periodic perturbations, with much lower associated signal on the orbits. There exists an exception, however, if one considers the special case of Lagrangian moons. Indeed, in such a case the tidal bulges are oriented on average with a constant angle close to 60° (see figure below).

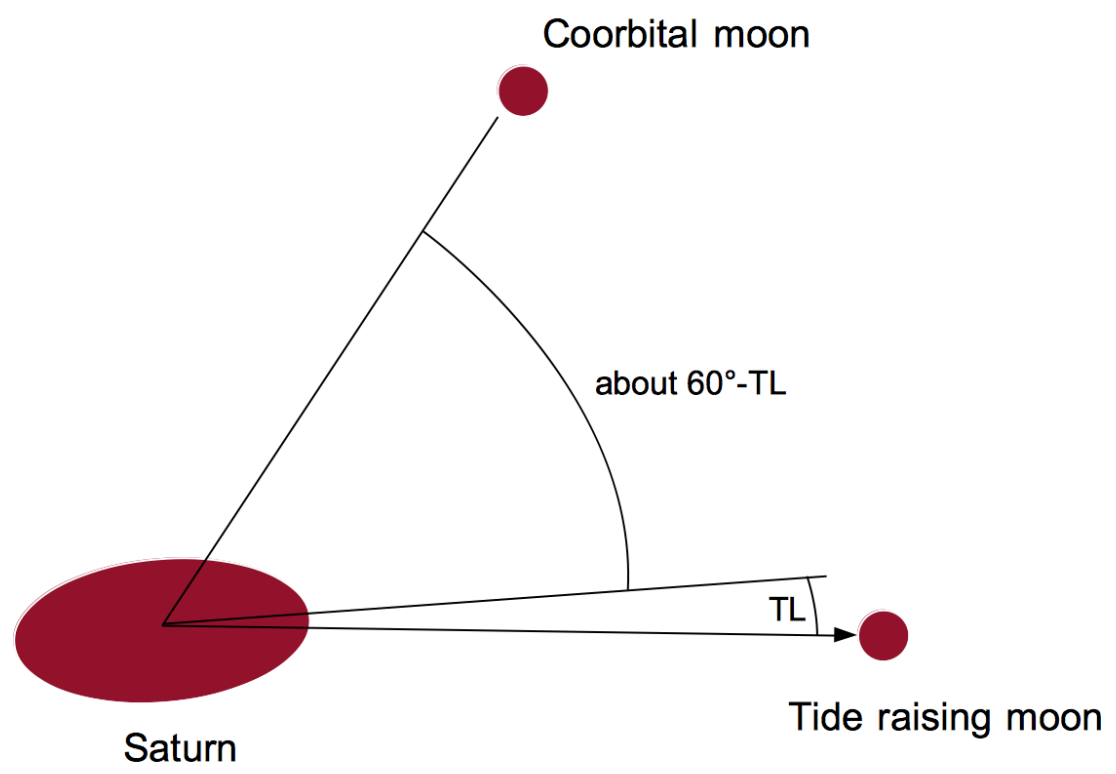

As a consequence, tidal effects arising on one moon and acting on a Lagrangian moon will provide a significant secular signature on the orbital longitude that is hopefully detectable. To quantify how large this effect can be, we rely here on numerical simulation. We provide below prefit and postfit residuals associated with these cross-tidal effects, for 14 moons of Saturn. The postfit simulations are obtained after having fitted all initial state vectors, masses, Saturn's $J_2$, polar orientation and precession, Saturn's tidal Q.

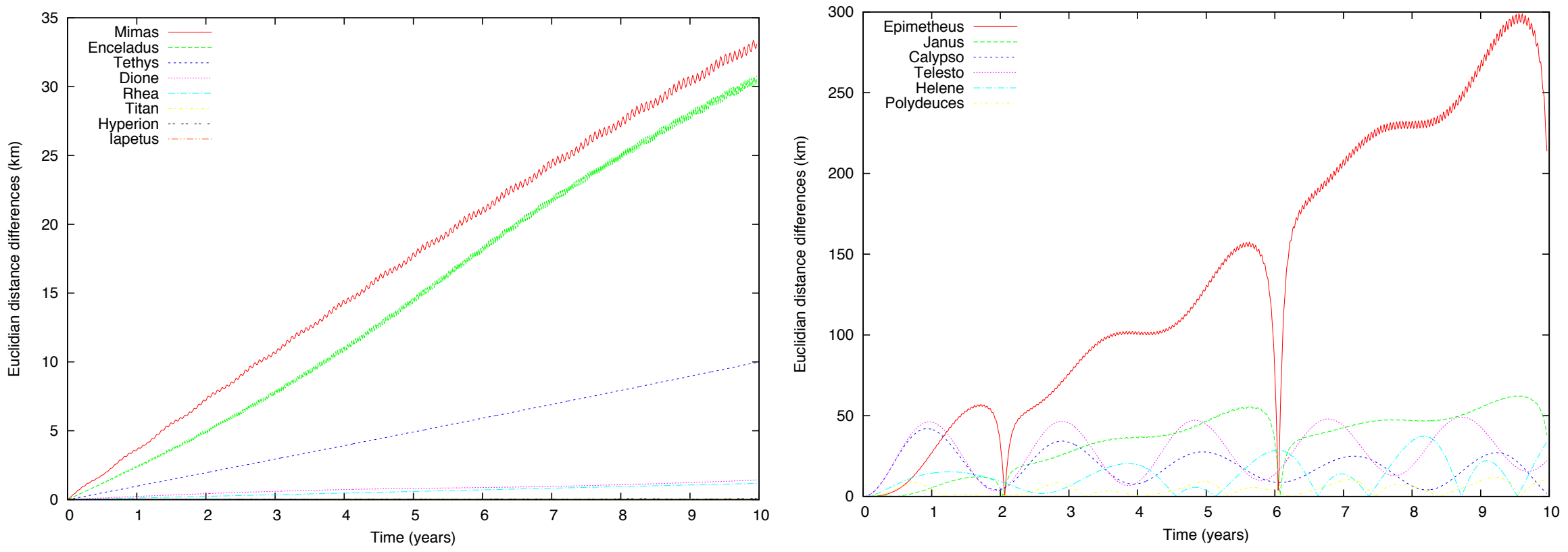

Figure A1.1: *Prefit residuals associated with cross-tidal effects.*

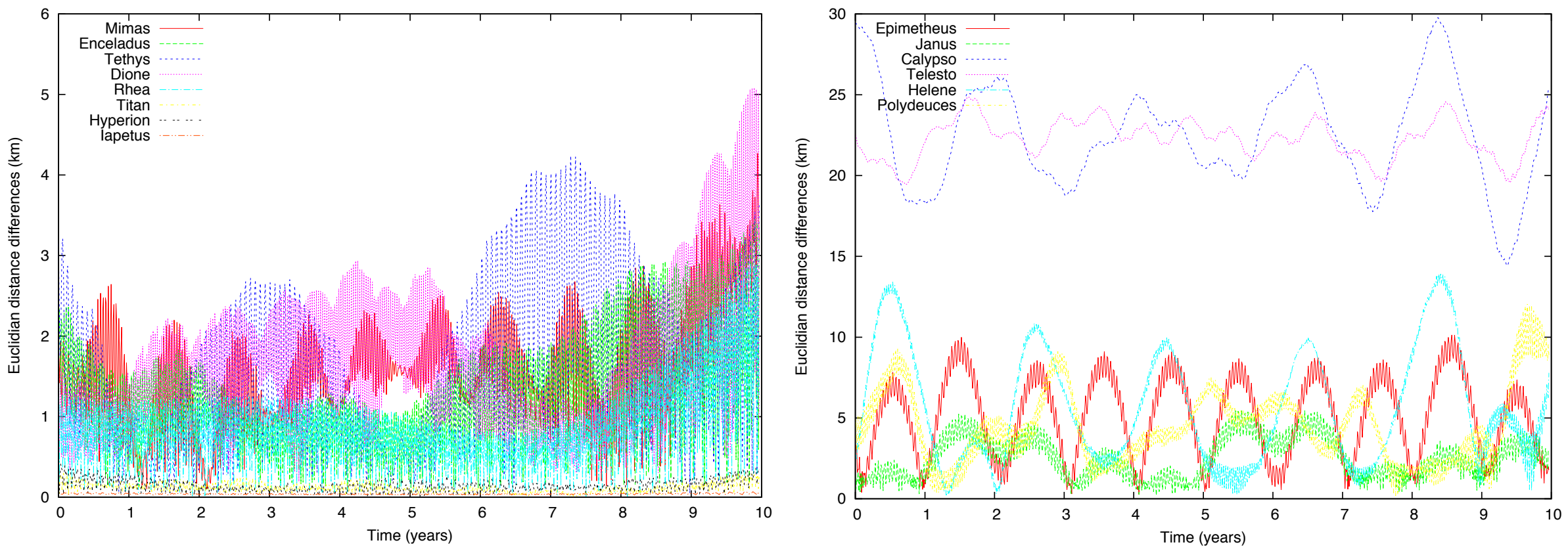

Figure A1.2: *Postfit residuals associated with cross-tidal effects.*

We can see that the largest effects indeed appear on the coorbital moons, with the highest effects on the Lagrangian satellites of Tethys and Dione. When not considering these cross-tidal effects, the astrometric residuals of these former moons can easily reach a few tens of kilometers, much above the typical 5 km residuals we obtained in the present work.

## A2 - Rhea's orbital acceleration under strong Saturnian tides

To estimate the impact of the large $k_2/Q$ value obtained at Rhea's tidal frequency, we perform prefit and postfit simulations (fitting the state vectors of all moons) over a century. Assuming $k_2/Q=122.28 \times 10^{-5}$ (see simulation 3 of ED.1), the postfit residuals below show that Rhea's longitude is affected by a signal of a bit more than 75 km.

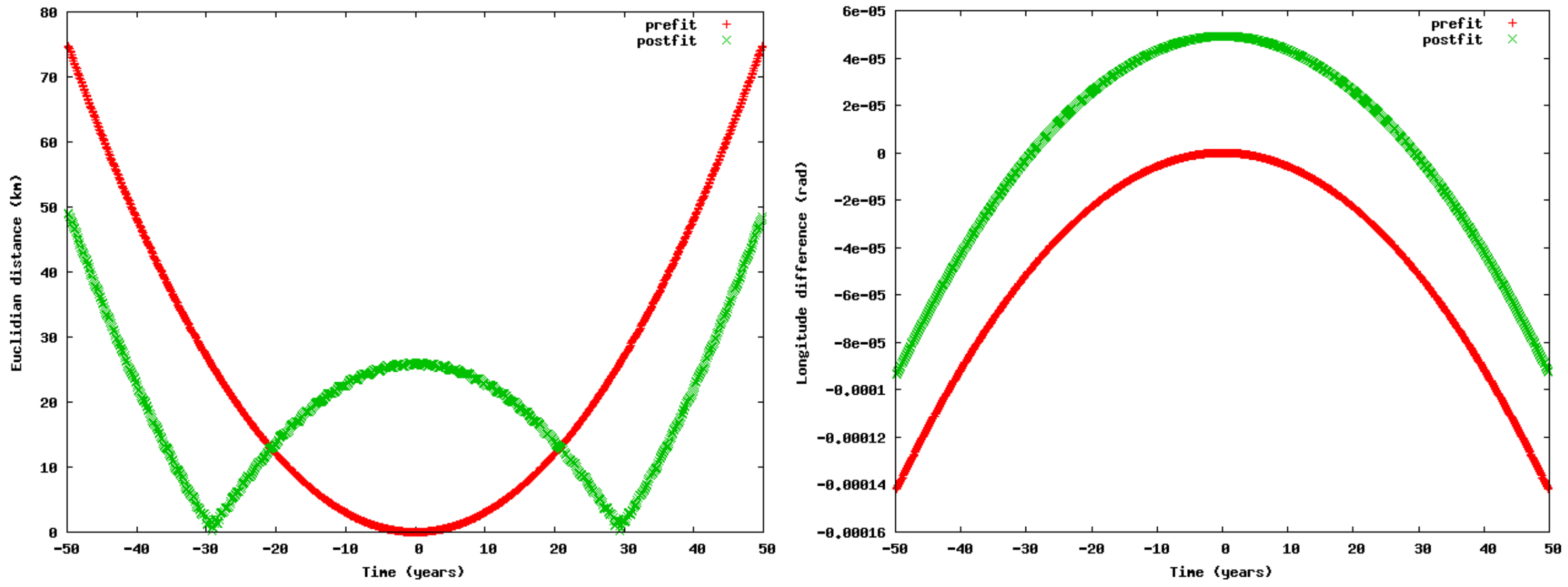

Figure A2.1: *Left: residuals in distance (km); right: residuals in the orbital longitude (rad)*

## A3 – Titan's possible past evolution from Saturn's low Q scenario

To investigate the effect of possible strong tides under Titan's orbital parameters, we can use analytical expression for da/dt and de/dt. In particular, limiting our study to Saturn and Titan, we recall that we have (as a first approximation) for the tides raised in the primary (Kaula, 1964):

$$\frac{da}{dt} = \frac{3k_2 mnR^5}{QMa^4}\left(1+\frac{51}{4}e^2\right)$$
$$\frac{de}{dt} = \frac{57k_2 mn}{8QM}\left(\frac{R}{a}\right)^5 e \qquad \text{(A1)}$$

and for the tides raised in the 1:1 spin-orbit satellite (Peale & Cassen, 1978):

$$\frac{da}{dt} = -\frac{21 k_2^s M n R_s^5}{Q^s m a^4} e^2$$
$$\frac{de}{dt} = -\frac{21 k_2^s M n}{2 Q^s m} \left( \frac{R_s}{a} \right)^5 e \qquad \text{(A2)}$$

the index *s* referring to the satellite.

To make the study straightforward, we first consider a two-body problem without tides inside Titan and assume no frequency dependence at all for Q. On Figure A3.1 we see that over the age of the Solar system, Titan's orbit decays pretty close to Saturn, but still above the Saturn's Roche limit. More, its eccentricity does progress from almost nil to its current value of about 0.03.

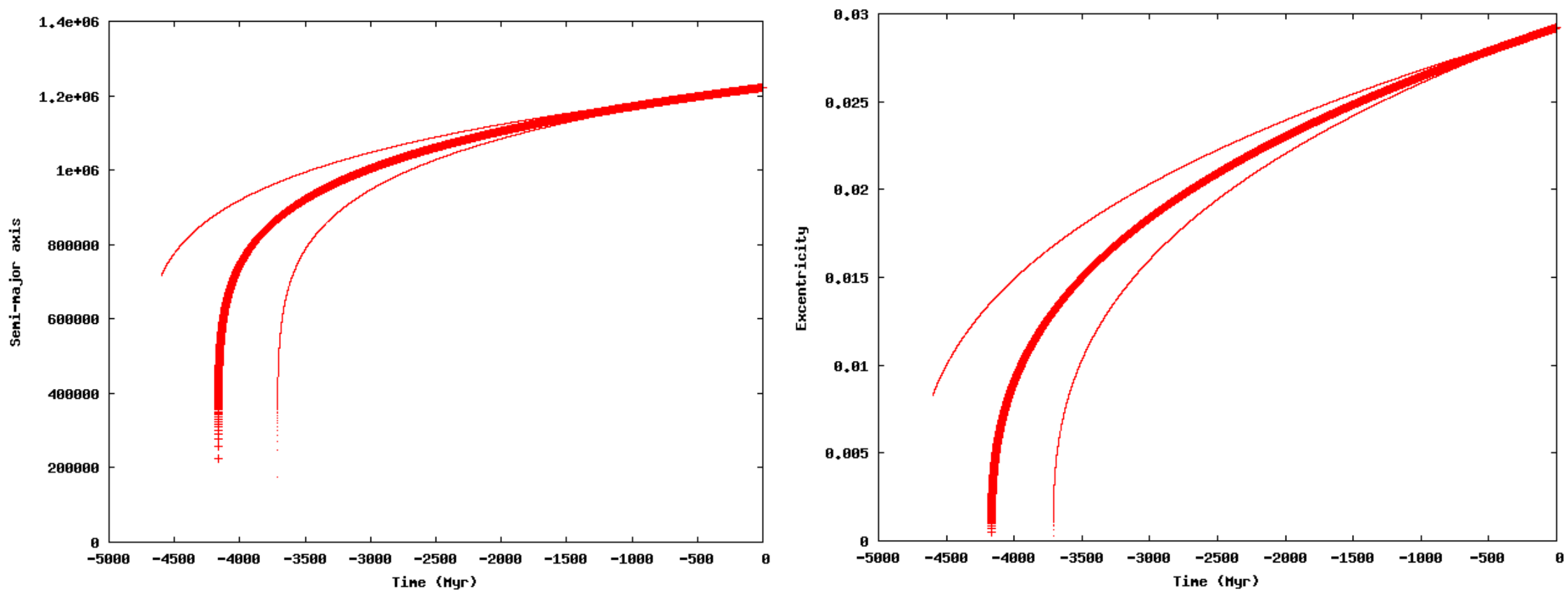

Figure A3.1: *Possible past evolution of Titan's semi-major axis and eccentricity assuming* $k_2/Q=(121.97 +/- 15.30) \times 10^{-5}$ *(merging of IMCCE's solutions from Tables 1-2) without tidal dissipation in the moon.*

In a second step, we start adding tidal dissipation with Titan. Here, only an average Q value is considered over 4.5 Byr, even though Q may have not been constant. We see form Figure A3.2 that low average value for Titan still allows Titan semi-major axis to evolve significantly, while its eccentricity can significantly change for the lowest dissipative solutions.

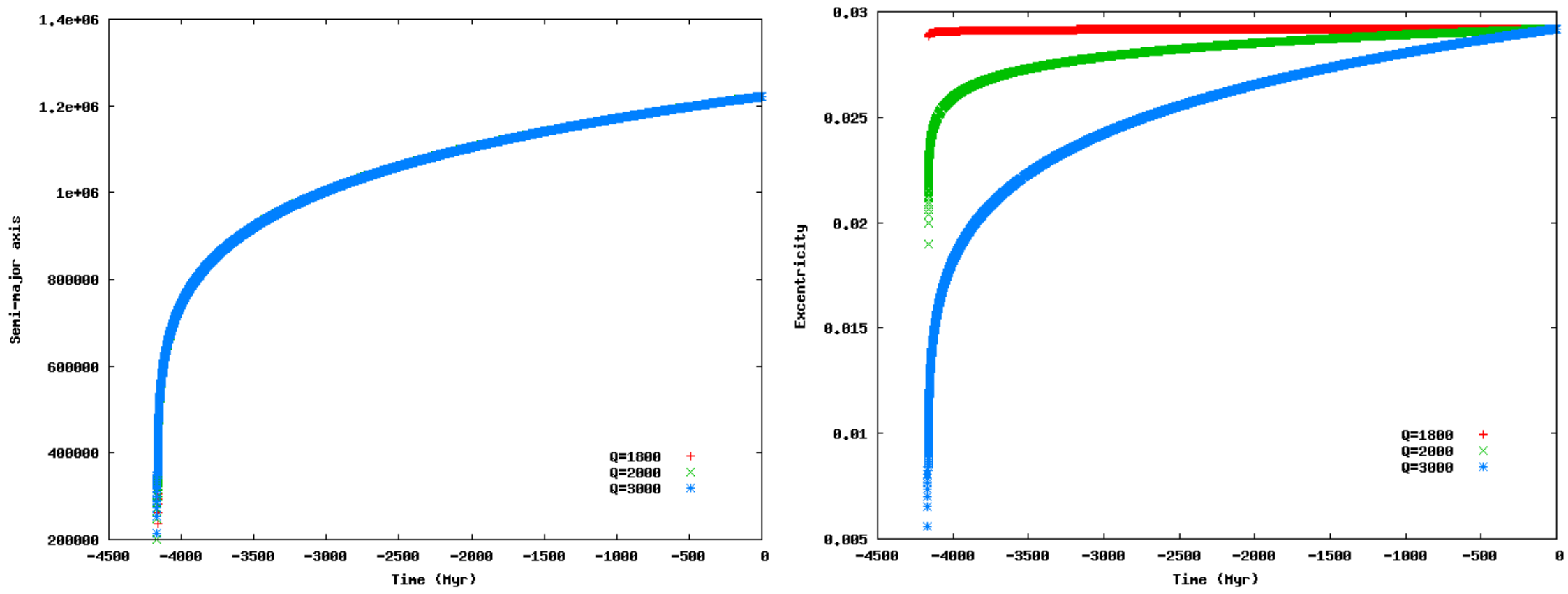

Figure A3.2: *Possible past evolution of Titan's semi-major axis and eccentricity assuming* $k_2/Q=(121.97 +/- 15.30) \times 10^{-5}$ *(merging of IMCCE's solutions from Tables 1-2), with tidal dissipation in the moon.*

We can conclude that the strong tidal scenario for Saturn may not in itself solve completely the question of the origin of Titan's high eccentricity, unless Titan has been poorly dissipative on average over the age of the Solar system. Another option might be that significant tidal dissipation occurs in Rhea also, offering a possible lower Q solution for Saturn. At least, we show that a higher dissipation in Saturn could be a key element in understanding the past evolution of Titan's formation and orbital evolution.

**A4 – Astrometric residuals and linear correlations**

To illustrate the various simulations that we performed, we provide astrometric residuals of the IMCCE solution that considered a constant $k_2/Q$ ratio and no tidal dissipation scenario within Enceladus. To save space, we do not provide here statistics of ground-based and HST data, since they are pretty similar to the ones published in Lainey et al. (2012). We provide below the plots of the O-Cs, only. Full statistics are available on request.

Figure A4.1 shows the astrometric residuals of the Lagrangian satellites of Tethys and Dione. Tables A4.1-4.3 provide the astrometric residuals of all observations for the 14 moons considered. Table A4.4 provides the correlations between all our fitted parameters and the tidal parameters $k_2$ and Q.

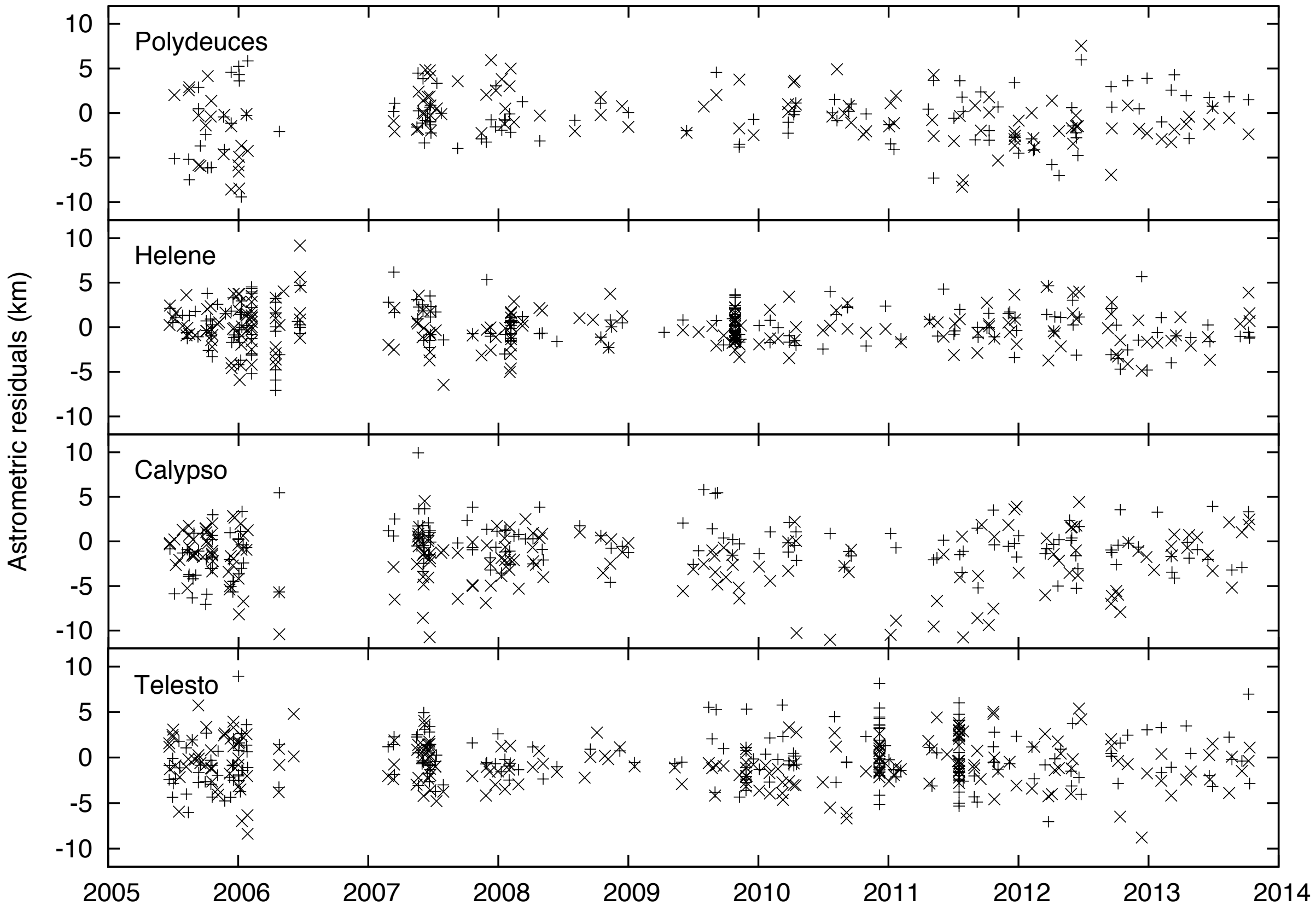

Figure A4.1: *Astrometric residuals of the four Lagrangian satellites from ISS-Cassini. Telesto and Calypso are the two coorbital moons of Tethys. They move around the Lagrangian stable points L4 and L5. Helene and Polydeuces are in equivalent orbital configurations but along the orbit of Dione. The associated ISS-NAC astrometric data are fitted in sample and line coordinates (pixel). Residuals are here converted to kilometres.*

| *Satellite* | $\mu_s$ | $\sigma_s$ | $\mu_l$ | $\sigma_l$ | $N_s$ | $N_l$ |
|---|---|---|---|---|---|---|
| Epimetheus | -0.0094 | 4.3180 | 0.1805 | 4.5340 | 350 | 350 |
| Janus | 0.0096 | 0.9780 | 0.5378 | 1.1566 | 322 | 322 |
| Mimas | 0.4190 | 0.2813 | -0.0460 | 0.6600 | 20 | 20 |
| Enceladus | -0.0014 | 0.3547 | -0.1116 | 0.2783 | 108 | 108 |
| Tethys | -0.1232 | 0.5284 | 0.0814 | 0.2600 | 25 | 25 |
| Dione | -0.0278 | 0.4808 | 0.0748 | 0.4730 | 84 | 84 |
| Rhea | -0.2925 | 0.4644 | -0.0035 | 0.2055 | 58 | 58 |
| Titan | 0.0000 | 0.0000 | 0.0000 | 0.0000 | 0 | 0 |
| Hyperion | 0.0000 | 0.0000 | 0.0000 | 0.0000 | 0 | 0 |
| Iapetus | 0.0000 | 0.0000 | 0.0000 | 0.0000 | 0 | 0 |
| Calypso | -0.0348 | 0.2508 | -0.1742 | 0.2546 | 230 | 230 |
| Telesto | -0.0190 | 0.2220 | -0.0366 | 0.2960 | 279 | 279 |
| Helene | -0.0164 | 0.2731 | -0.0456 | 0.2492 | 262 | 262 |
| Polydeuces | -0.0554 | 0.2508 | -0.0584 | 0.2422 | 139 | 139 |

Table A4.1 (one single moon per image): *Statistics of the ISS-NAC astrometric residuals computed from IMCCE model (no tidal dissipation within Enceladus scenario) in pixel. $\mu$ and $\sigma$ denote respectively the mean and standard deviation of the residuals computed on sample and line. $N_s$ and $N_l$ are the number of observations considered for the respective coordinate.*

| *Satellite* | $\mu_s$ | $\sigma_s$ | $\mu_l$ | $\sigma_l$ | $N_s$ | $N_l$ |
|---|---|---|---|---|---|---|
| Epimetheus | 0.0203 | 0.2778 | 0.0449 | 0.2912 | 28 | 28 |
| Janus | -0.0203 | 0.2778 | -0.0449 | 0.2912 | 28 | 28 |
| Mimas | 0.0255 | 0.1784 | -0.0064 | 0.2745 | 134 | 134 |
| Enceladus | -0.0307 | 0.1784 | 0.0084 | 0.1248 | 327 | 327 |
| Tethys | 0.0211 | 0.1088 | 0.0186 | 0.1359 | 424 | 424 |
| Dione | -0.0204 | 0.1061 | 0.0054 | 0.1070 | 592 | 592 |
| Rhea | 0.0175 | 0.1370 | -0.0234 | 0.1208 | 556 | 556 |
| Titan | 0.0000 | 0.0000 | 0.0000 | 0.0000 | 0 | 0 |
| Hyperion | 0.0000 | 0.0000 | 0.0000 | 0.0000 | 0 | 0 |
| Iapetus | 0.0000 | 0.0000 | 0.0000 | 0.0000 | 0 | 0 |
| Calypso | 0.1470 | 0.0000 | -0.5137 | 0.0000 | 1 | 1 |
| Telesto | -0.0997 | 0.0702 | 0.2454 | 0.1691 | 3 | 3 |
| Helene | -0.1308 | 0.0508 | 0.2090 | 0.0096 | 2 | 2 |
| Polydeuces | 0.1379 | 0.0731 | -0.2135 | 0.1657 | 3 | 3 |

Table A4.2 (multiple moon per image): *Statistics of the ISS-NAC astrometric residuals computed from IMCCE model (no tidal dissipation within Enceladus scenario) in pixel. μ and σ denote respectively the mean and standard deviation of the residuals computed on sample and line.* $N_s$ *and* $N_l$ *are the number of observations considered for the respective coordinate.*

| *Satellite* | $\mu_{RA}$ | $\sigma_{RA}$ | $\mu_{DEC}$ | $\sigma_{DEC}$ | $N_{RA}$ | $N_{DEC}$ |
|---|---|---|---|---|---|---|
| Mimas | -1.1001 | 3.9151 | -1.1401 | 2.8370 | 826 | 826 |
| Enceladus | -0.1979 | 2.8234 | 0.2713 | 2.6588 | 732 | 732 |
| Tethys | 0.0532 | 4.5654 | -0.0123 | 3.5007 | 924 | 924 |
| Dione | -0.2068 | 4.1726 | -0.5264 | 3.4948 | 948 | 949 |
| Rhea | -0.3170 | 3.3581 | -0.1138 | 2.4739 | 1021 | 1021 |
| Titan | 0.0000 | 0.0000 | 0.0000 | 0.0000 | 0 | 0 |
| Hyperion | -0.1292 | 15.4526 | -5.9373 | 12.7287 | 92 | 90 |
| Iapetus | 1.4754 | 5.1951 | -1.1544 | 5.4322 | 1534 | 1534 |

Table A4.3 (one moon per image): *Statistics of the ISS-NAC astrometric residuals computed from IMCCE model (no tidal dissipation within Enceladus scenario) in km. μ and σ denote respectively the mean and standard deviation of the residuals computed on RA and DEC.* $N_{RA}$ *and* $N_{DEC}$ *are the number of observations considered for the respective coordinate.*

| | $k_2$ | Q |
|---|---|---|
| $a_1$ | 0.006 | 0.023 |
| $l_1$ | 0.002 | -0.014 |
| $k_1$ | -0.000 | -0.001 |
| $h_1$ | 0.002 | 0.002 |
| $q_1$ | -0.000 | -0.002 |
| $p_1$ | 0.000 | 0.003 |
| $a_2$ | 0.008 | 0.025 |
| $l_2$ | -0.004 | -0.029 |
| $k_2$ | -0.001 | 0.002 |
| $h_2$ | -0.002 | 0.001 |
| $q_2$ | 0.000 | -0.001 |
| $p_2$ | -0.000 | 0.002 |
| $a_3$ | 0.009 | 0.025 |
| $l_3$ | -0.013 | 0.232 |
| $k_3$ | -0.013 | 0.017 |
| $h_3$ | -0.003 | 0.002 |
| $q_3$ | 0.017 | -0.024 |
| $p_3$ | 0.002 | 0.070 |
| $a_4$ | 0.009 | 0.027 |
| $l_4$ | -0.012 | 0.182 |
| $k_4$ | 0.017 | 0.084 |
| $h_4$ | -0.026 | -0.026 |
| $q_4$ | 0.004 | -0.000 |
| $p_4$ | -0.006 | 0.127 |
| $a_5$ | 0.009 | 0.024 |
| $l_5$ | 0.009 | -0.223 |
| $k_5$ | 0.000 | 0.020 |
| $h_5$ | -0.003 | -0.074 |
| $q_5$ | -0.027 | 0.012 |
| $p_5$ | 0.011 | 0.069 |
| $a_6$ | 0.009 | 0.026 |
| $l_6$ | 0.002 | -0.509 |
| $k_6$ | 0.011 | -0.005 |
| $h_6$ | -0.010 | 0.082 |
| $q_6$ | 0.005 | -0.012 |
| $p_6$ | -0.007 | 0.154 |
| $a_7$ | 0.009 | 0.023 |
| $l_7$ | -0.003 | -0.216 |
| $k_7$ | -0.006 | -0.029 |
| $h_7$ | -0.003 | -0.008 |
| $q_7$ | -0.006 | 0.203 |
| $p_7$ | -0.007 | 0.036 |
| $a_8$ | 0.010 | 0.019 |
| $l_8$ | -0.002 | -0.005 |
| $k_8$ | -0.002 | -0.003 |
| $h_8$ | 0.003 | 0.025 |
| $q_8$ | 0.006 | 0.059 |
| $p_8$ | 0.002 | -0.013 |
| $a_9$ | 0.007 | 0.016 |
| $l_9$ | -0.001 | -0.005 |
| $k_9$ | -0.001 | 0.001 |
| $h_9$ | 0.002 | 0.014 |
| $q_9$ | -0.003 | -0.000 |
| $p_9$ | 0.000 | -0.018 |
| $a_{10}$ | 0.008 | 0.008 |
| $l_{10}$ | -0.004 | -0.007 |
| $k_{10}$ | -0.008 | -0.005 |
| $h_{10}$ | -0.007 | -0.007 |
| $q_{10}$ | 0.000 | 0.005 |
| $p_{10}$ | -0.002 | -0.022 |
| $a_{11}$ | 0.010 | 0.025 |
| $l_{11}$ | -0.024 | -0.114 |
| $k_{11}$ | 0.034 | 0.003 |
| $h_{11}$ | -0.012 | -0.002 |
| $q_{11}$ | -0.028 | 0.029 |
| $p_{11}$ | 0.018 | 0.051 |
| $a_{12}$ | 0.008 | 0.025 |
| $l_{12}$ | 0.142 | -0.216 |
| $k_{12}$ | -0.002 | -0.011 |
| $h_{12}$ | -0.012 | -0.006 |
| $q_{12}$ | 0.025 | -0.018 |
| $p_{12}$ | 0.011 | 0.026 |
| $a_{13}$ | 0.005 | 0.025 |
| $l_{13}$ | -0.028 | -0.254 |
| $k_{13}$ | 0.010 | 0.033 |
| $h_{13}$ | -0.002 | 0.026 |
| $q_{13}$ | -0.000 | -0.031 |
| $p_{13}$ | 0.001 | 0.062 |
| $a_{14}$ | 0.010 | 0.029 |
| $l_{14}$ | -0.073 | -0.254 |
| $k_{14}$ | 0.020 | -0.055 |
| $h_{14}$ | 0.007 | -0.052 |
| $q_{14}$ | 0.004 | -0.021 |
| $p_{14}$ | -0.005 | 0.054 |
| M | 0.009 | 0.026 |
| $m_1$ | -0.004 | 0.003 |
| $m_2$ | -0.004 | 0.003 |
| $m_3$ | -0.001 | -0.378 |
| $m_4$ | 0.038 | -0.064 |
| $m_5$ | 0.118 | -0.019 |
| $m_6$ | 0.120 | 0.029 |
| $m_7$ | 0.011 | -0.062 |
| $m_8$ | 0.000 | 0.004 |
| $m_9$ | 0.000 | -0.003 |
| $m_{10}$ | -0.005 | -0.011 |
| $a_0$ | 0.003 | -0.591 |
| $d_0$ | -0.010 | 0.138 |
| $c_{20}$ | -0.005 | 0.014 |
| da/dt | 0.017 | 0.186 |
| dd/dt | 0.012 | -0.129 |
| $k_2$ | 1.000 | -0.030 |
| Q | -0.030 | 1.000 |

Table A4.4: *Correlation between all our fitted parameters and the tidal parameters $k_2$ and Q. Here a is the semi-major axis, l is the mean longitude, e is the eccentricity, Ω is the longitude of the node, ω is the argument of the periapsis, k=e cos(Ω+ω), h=e sin(Ω+ω), q=sin(i/2) cos(Ω) and p=sin(i/2) sin(Ω). Numbers 1,2,3…14 refer to Epimetheus, Janus, the eight main moons (Mimas,...Iapetus), Calypso, Telesto, Helene, Polydeuces, respectively. Full table is available on request.*

**Acknowledgements**: The authors are indebted to all participants of the Encelade WG. V.L. would like to thank Michael Efroimsky for fruitful discussions. This work has been supported by the European Community's Seventh Framework Program (FP7/2007-2013) under grant agreement 263466 for the FP7-ESPaCE project, the International Space Science Institute (ISSI), PNP (INSU/CNES) and AS GRAM (INSU/CNES/INP). The work of R. A. J. was carried out at the Jet Propulsion Laboratory, California Institute of Technology, under a contract with NASA. N.C. and C.M. were supported by the UK Science and Technology Facilities Council (Grant No. ST/M001202/1) and are grateful to them for financial assistance. C.M. is also grateful to the Leverhulme Trust for the award of a Research Fellowship. N.C. thanks the Scientific Council of the Paris Observatory for funding. S. Mathis acknowledge funding by the European Research Council through ERC grant SPIRE 647383. The authors are indebted to the Cassini project and the Imaging Science Subsystem Team for making this collaboration possible.